\documentclass[12pt]{article}
\usepackage{amsmath}
\usepackage{graphicx,psfrag,epsf}
\usepackage{enumerate}
\usepackage{natbib}
\bibpunct{(}{)}{;}{a}{}{,}
\usepackage{url} 
\providecommand{\harvardurl}{}\renewcommand{\harvardurl}[1]{\textbf{URL:} \url{#1}}

\newcommand{\blind}{0}

\usepackage{multirow}
\usepackage{makecell}
\usepackage{booktabs}
\usepackage[usenames,dvipsnames]{xcolor}
\usepackage{amssymb}
\usepackage{amsfonts}
\usepackage{mathrsfs} 
\usepackage{physics}
\usepackage{mathtools}
\usepackage{algorithm}
\usepackage{algpseudocode}
\usepackage{caption}
\usepackage{subcaption}
\usepackage{amsthm}
\newtheorem{proposition}{Proposition}
\newtheorem{lemma}{Lemma}
\newtheorem{theorem}{Theorem}

\theoremstyle{remark}
\newtheorem{remark}{Remark}
\newcommand{\bs}{\boldsymbol}
\DeclareMathOperator*{\argmin}{arg\,min}
\usepackage{hyperref}

\newcommand{\ra}[1]{\renewcommand{\arraystretch}{#1}}
\begin{document}

\def\spacingset#1{\renewcommand{\baselinestretch}%
{#1}\small\normalsize} \spacingset{1}


\if0\blind
{
  \title{\bf Clouded Representatives and Score Matching for Distributed Quantile Regression}
  \author{Keren Li\hspace{.2cm}\\
    Department of Mathematics, University of Alabama at Birmingham\\
    and \\
    Boxiang Wang \\
    Department of Statistics and Actuarial Science, University of Iowa}
  \maketitle
} \fi

\if1\blind
{
  \bigskip
  \bigskip
  \bigskip
  \begin{center}
    {\LARGE\bf Clouded Representatives and Score Matching for Distributed Quantile Regression}
\end{center}
  \medskip
} \fi

\bigskip
\begin{abstract}
Quantile regression is widely used in high-stakes applications where the data are distributed across many sites and the individual records cannot be shared. The existing distributed methods for quantile regression typically require homogeneous sites, large sites, or synchronous communication, which are rarely all satisfied in practice. In this paper, we propose a new method named cloudqr. Each site summarizes its observations by a few Gaussian clouds, each given by the size, center, and covariance of a group of observations, such that the clouds match the score of the raw data exactly. A central server then fits the quantile regression on the clouds from all sites. We establish the convergence of cloudqr under mild conditions. With simulation studies and two real-world applications, we demonstrate the superior performance of cloudqr over the existing distributed methods.
\end{abstract}

\noindent%
\vfill

\newpage
\spacingset{1.45} 

\section{Introduction}\label{sec:intro}

Quantile regression \citep{koenker1978regression} has been a fundamental tool in statistics for nearly fifty years. By modeling the quantiles of a response given the covariates at different levels, it provides a comprehensive picture of the conditional distribution. In particular, quantile regression characterizes the lower and upper tails, which are often of primary interest in high-stakes applications. For example, quantile regression is used in medicine to construct reference ranges such as growth charts \citep{wei2006quantile} and the age-specific percentiles in the benchmark GAGurine data \citep{venables2002mass, tang2026fastkqr}, in public health to study low birth weight \citep{koenker2001quantile,abrevaya2001effects} and survival times \citep{peng2021quantile}, and in finance to estimate the value at risk of a portfolio \citep{engle2004caviar} and the systemic risk of a financial institution \citep{adrian2016covar}. Readers are referred to \citet{koenker2005quantile,koenker2017forty} and \citet{koenker2017handbook} for more applications and developments. In these high-stakes applications, accurate estimation is essential, and ideally all the available data would be pooled. However, such data usually come from different hospitals, insurers, or banks, and they are confidential, so moving individual records to one place is typically not allowed \citep{kairouz2021advances,tong2022distributed}. The data therefore have to be analyzed in a distributed setting, where the records stay at the sites and only summaries, such as counts and averages, can be communicated.

In this paper, we study quantile regression in such a distributed setting, with the goal of estimating its coefficients as if the data from all sites were pooled together. In applications such as medicine and finance, three features of the distributed setting make this goal challenging. First, the sites are often heterogeneous, as they serve different populations under different practices \citep{tong2022distributed}. These differences can be much larger in the tails than at the center: across 37 children's hospitals treating asthma, the median length of stay was one or two days at every hospital, whereas the 90th percentile ranged from two to five days \citep{silber2016auditing}. Second, some sites can be very small: a clinical network may have hundreds of sites, some of which contribute only a few patients \citep{tong2022distributed,duan2022heterogeneity}. Third, the sites often reply on their own schedules, so some replies arrive late \citep{bonawitz2019towards,kairouz2021advances}. A method for distributed quantile regression should accommodate all three features.

Although distributed estimation has been extensively studied in statistics \citep{zhang2013communication,battey2018distributed,jordan2019communication,fan2023communication} and in machine learning under the name of federated learning \citep{mcmahan2017communication,kairouz2021advances}, much less work has been done on distributed quantile regression. In the literature, \citet{volgushev2019distributed} averaged the local fits of the sites, assuming that every site is large enough for a local fit and that the sites are homogeneous. \citet{chen2019quantile} proposed a divide-and-conquer estimator that starts from an initial estimate computed at one site and successively refines it with summaries from all sites; see also \citet{chen2020quantile} and \citet{chen2020distributed}. In \citet{jordan2019communication}, \citet{tan2022communication}, and \citet{yang2026iterative}, one site serves as the \textit{master}: it collects summaries from the other sites and solves the problem on its own data with the help of these summaries. Their theory requires the objective at the master to be close to the pooled objective. However, a master site can rarely be designated in these applications: a hospital or a bank does not serve as the master for its peers, and the standard practice is to coordinate the analysis through a \textit{central server} that holds no raw data \citep{kairouz2021advances,rieke2020future}.

To accommodate heterogeneous sites, small sites, and late replies, we propose a new method, dubbed cloudqr, for distributed quantile regression. This method is motivated by representative learning \citep{li2022smr,zheng2024rasmr,li2026asmr}, in which each site summarizes its own data by \textit{representatives}, and a central server fits the model on those representatives instead of the raw data. The existing methods craft pseudo observations as representatives, whose scores match those of their groups for smooth losses. The quantile fit of a group also depends on which side of the fit its observations lie on, not only on their center. Therefore, we treat each observation as a small Gaussian cloud, and the fraction of the cloud on either side measures the tendency of the observation to lie above or below the fit. Averaging the check loss over each cloud also amounts to an implicit smoothing of the loss, which coincides with the convolution-smoothed check loss studied by \citet{fernandes2021smoothing} and \citet{he2023smoothed}. We show that the clouds in a group can be combined into one Gaussian cloud with the same score, which summarizes the group by its size, center, and covariance; we call it a \textit{clouded representative}. Our cloudqr method proceeds in rounds. In each round, each site calculates its own clouded representatives at the current quantile regression fit, and the server updates the fit using all of them as if they were the data; the rounds repeat until convergence. Neither a local fit nor a master site is needed. Moreover, the server does not need to wait for late sites, because it can keep updating the fit with the representatives already received. It is also worth emphasizing that summaries of this kind are routinely reported in medical studies \citep{vonelm2007strobe,hripcsak2015ohdsi}. Unlike a gradient or a Hessian matrix that other distributed methods communicate, a clouded representative can be read and checked, so the communication is interpretable. This is valued in high-stakes applications \citep{rudin2019stop,tonekaboni2019clinicians}.

\begin{figure}[!t]
\centering
\includegraphics[width=0.9\textwidth]{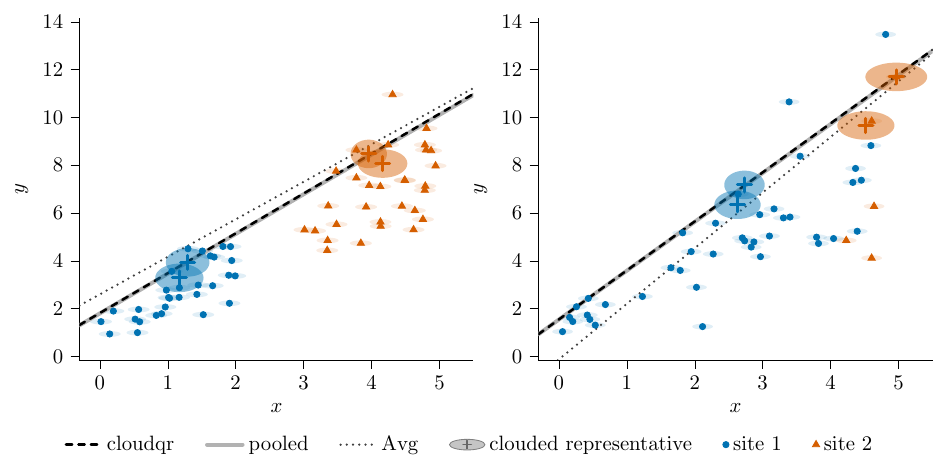}
\caption{Two toy examples of distributed quantile regression at the level $\tau = 0.9$. In both panels the data are generated from $Y = 1 + 1.5X + (0.4 + 0.3X)\varepsilon$, where $\varepsilon \sim \mathcal N(0,1)$. In the left panel each site has 30 observations, with covariates on $[0, 2]$ and on $[3, 5]$; in the right panel the two sites have 40 and 5 observations. The light gray line is the quantile regression fit on the pooled data, the dotted line is the average of the site-wise fits (Avg), and the dashed line is the cloudqr fit. The small clouds are the clouded observations, and the larger clouds marked by a plus sign are the clouded representatives of the final round, with widths proportional to their standard deviations.}
\label{fig:toy}
\end{figure}

To give a quick illustration, Figure~\ref{fig:toy} shows two toy examples with two sites at the quantile level $\tau=0.9$: heterogeneous sites in the left panel, whose covariates lie on different ranges, and a small site in the right panel, with only five observations. In both cases, the average of the two site-wise fits is far from the fit on the pooled data, whereas our cloudqr fit almost overlaps with it. Each observation is lifted to a small Gaussian cloud, and each site combines its clouds into two larger clouds, say the clouded representatives, which are the summaries that each site sends to the central server.

We construct the clouded representatives in closed form and show that their score exactly matches the score of the raw observations lifted to their clouds. We establish the convergence of cloudqr when the representatives are refreshed at every step and bound its error when they are reused or arrive late. Extensive simulation studies support our theoretical findings and show that cloudqr closely matches the pooled fit in settings where the existing methods fail. We further demonstrate cloudqr in two high-stakes real-world applications, the length of stay of patients across hospitals and the interest rate of small business loans across lenders. We implement cloudqr in an R package of the same name.

The rest of the paper is organized as follows. We construct the clouded representatives in Section~\ref{sec:clouded_rep} and establish their theoretical properties in Section~\ref{sec:theory}. Simulation studies and the real data analysis are given in Sections~\ref{sec:simu} and~\ref{sec:data}. Section~\ref{sec:diss} discusses extensions.

\section{Clouded Representatives}\label{sec:clouded_rep}

In this section, we develop cloudqr for distributed quantile regression. We first review quantile regression and specify the distributed setting in Section~\ref{sec:setup}, then lift the observations to Gaussian clouds in Section~\ref{sec:clouds}, construct the clouded representatives in Section~\ref{sec:cloud_smr}, and present the algorithm in Section~\ref{sec:algorithm}.
\subsection{Background and Setup}\label{sec:setup}

Consider first the standard setting, in which all data are stored at one place. Suppose that the training data consist of $n$ observations $\{(y_i,\mathbf x_i)\}_{i=1}^n$, where $\mathbf x_i\in\mathbb R^p$ and $y_i\in\mathbb R$. For a given level $\tau\in(0,1)$, let $Q_\tau(y\mid\mathbf x)$ denote the $\tau$th conditional quantile of $y$ given $\mathbf x$, that is, $Q_\tau(y\mid\mathbf x)=\inf\{q:\mathrm{P}(y\le q\mid\mathbf x)\ge\tau\}$. The linear quantile regression model assumes that $Q_\tau(y\mid\mathbf x)=\beta_0+\mathbf x^\top\bs\beta$, where $\beta_0\in\mathbb R$ and $\bs\beta\in\mathbb R^p$ are the quantile regression coefficients. For simplicity, we write $\bs\theta=(\bs\beta,\beta_0)$. To estimate $\bs\theta$, linear quantile regression solves
\begin{align}\label{eq:pooled_qr}
\hat{\bs\theta}_{\mathrm{QR}}=\argmin_{\bs\theta}\left\{\frac1n\sum_{i=1}^n\rho_\tau(y_i-\mathbf x_i^\top\bs\beta-\beta_0)\right\},
\end{align}
where $\rho_\tau(t)=t\{\tau-\mathbb{I}(t<0)\}$ is the check loss.

Suppose now that the $n$ observations are stored separately at $K$ sites. For each $k=1,2,\ldots,K$, we define $\mathscr D_k=\{(y_i,\mathbf x_i): i\in I_k\}$ to collect the $n_k$ observations at site $k$, and the index sets $I_1,\ldots,I_K$ partition $\{1,\ldots,n\}$. The fit $\hat{\bs\theta}_{\mathrm{QR}}$ in \eqref{eq:pooled_qr} is obtained using all $n$ observations at one place, so we call it the pooled fit and use it as the benchmark.

Following the standard setting of representative learning \citep{li2022smr,zheng2024rasmr,li2026asmr}, we consider the situation where summaries of the observations, such as counts and averages, are permitted to be shared. Specifically, site $k$ partitions its observations into $m_k$ groups, with index sets $I_{k1},\ldots,I_{km_k}$, and summarizes each group by a \textit{point representative}, so that site $k$ sends the representative data set $\check{\mathscr D}_k = \{ (n_{kh}, \check y_{kh}, \check{\mathbf x}_{kh}) \}_{h=1}^{m_k}$ in place of $\mathscr D_k$. At site $k$, the $h$th representative consists of the group size $n_{kh}=|I_{kh}|$, a response $\check y_{kh}$, and a covariate $\check{\mathbf x}_{kh}$, so each representative keeps the size and the center of its group. The server fits the model on $\check{\mathscr D}_1, \check{\mathscr D}_2, \ldots, \check{\mathscr D}_K$.

\subsection{Gaussian Clouds}\label{sec:clouds}

A point representative in $\check{\mathscr D}_k$ has a single residual, with one sign, but the check loss of a group depends on the signs of all its residuals. To capture such information, we propose to lift each observation to a Gaussian cloud, $\mathbf X_i\sim\mathcal N(\mathbf x_i,\, s^2 \mathbf{I}_p)$, where $s>0$ is the cloud scale. Let $Q_n$ denote the collection of the $n$ clouds. The clouded risk is the average of the check loss over the clouds,
\begin{align}\label{eq:lifted_quantile_risk}
 R_{Q_n}(\bs\theta)
 =
 \frac{1}{n}\sum_{i=1}^n
 \mathrm{E}_{\mathbf{X}_i\sim \mathcal N(\mathbf x_i,\, s^2 \mathbf{I}_p)}
 \Bigl[
 \rho_\tau\bigl(y_i-\mathbf{X}_i^\top \bs\beta- \beta_0\bigr)
 \Bigr].
\end{align}
The random residual $y_i-\mathbf{X}_i^\top \bs\beta- \beta_0$ follows a univariate normal distribution $\mathcal N(r_i, s_{\bs\beta}^2)$, in which $r_i(\bs\theta) = y_i-\mathbf{x}_i^\top\bs\beta-\beta_0$ is deterministic and $s_{\bs\beta}=s\|\bs\beta\|$, where $\|\cdot\|$ is the Euclidean norm. We assume $\bs\beta\neq\mathbf 0$ throughout.
Writing the random residual as $r_i + s_{\bs\beta} Z$ with $Z \sim \mathcal N(0,1)$, we obtain the clouded check loss
\begin{align}\label{eq:expected_check_loss}
 \mathrm{E}\bigl[\rho_\tau(y_i-\mathbf{X}_i^\top \bs\beta- \beta_0)\bigr]
 = r_i(\bs\theta)\, w_i(\bs\theta) + s_{\bs\beta}\,\phi\bigl(a_i(\bs\theta)\bigr),
\end{align}
where $a_i(\bs\theta)=r_i(\bs\theta)/s_{\bs\beta}$, $w_i(\bs\theta)=\tau-\Phi\bigl(-a_i(\bs\theta)\bigr)$, and $\Phi$ and $\phi$ denote the cumulative distribution function and the density function of the standard normal distribution, respectively.

In the sequel, we suppress the argument $\bs\theta$ when no confusion arises. Substituting \eqref{eq:expected_check_loss} into \eqref{eq:lifted_quantile_risk} yields $R_{Q_n}(\bs\theta) = \frac{1}{n}\sum_{i=1}^n \{ r_i w_i + s_{\bs\beta}\,\phi(a_i) \}$.
As $R_{Q_n}(\bs\theta)$ is a smooth function of
$\bs\theta$, direct differentiation gives its gradient,
\begin{align}\label{eq:grad_lifted_quantile_risk_gaussian}
\nabla_{\bs\beta} R_{Q_n}= \frac{1}{n}\sum_{i=1}^n\left\{-\mathbf{x}_i w_i+ \phi\bigl(a_i\bigr)\frac{s_{\bs\beta} \bs\beta}{\|\bs\beta\|^2} \right\},\qquad
\nabla_{\beta_0} R_{Q_n}= -\frac{1}{n}\sum_{i=1}^n w_i.
\end{align}
Following the score-matching representatives of \citet{li2022smr}, we also call the gradient the score, and we use the two terms interchangeably.
\begin{remark}\label{rem:smoothing}
For the squared loss, averaging over the clouds gives the least-squares loss plus the ridge penalty $s^2\|\bs\beta\|^2$, since $\mathrm{E}\{(y-\mathbf X^\top\bs\beta-\beta_0)^2\}=(y-\mathbf x^\top\bs\beta-\beta_0)^2+s^2\|\bs\beta\|^2$ for $\mathbf X\sim\mathcal N(\mathbf x,s^2\mathbf{I}_p)$.
\end{remark}

\begin{remark}\label{rem:smoothing2}
The clouded check loss has a neat connection to the density-convoluted quantile regression of \citet{fernandes2021smoothing} and \citet{he2023smoothed}. Since $\mathbf X_i^\top\bs\beta$ is normal with standard deviation $s\|\bs\beta\|$, the clouded check loss \eqref{eq:expected_check_loss} coincides with the check loss in \eqref{eq:pooled_qr} convolved with a Gaussian kernel of bandwidth $s\|\bs\beta\|$. The density convolution turns the check loss into a smooth objective, whereas our clouded check loss is designed to capture the signs of the residuals so that a group can be summarized by one representative.
\end{remark}

\subsection{Clouded Representatives}\label{sec:cloud_smr}

With observations lifted to Gaussian clouds, we can now construct clouded representatives. Let site $k$ have groups $I_{k1},\ldots,I_{km_k}$. For each group, $h=1,\ldots,m_k$, we add a new covariance term $\tilde{\bs\Sigma}_{kh}$ to the representatives. In particular, the clouded representative of group $h$ is $(n_{kh},\tilde y_{kh},\tilde{\mathbf x}_{kh},\tilde{\bs\Sigma}_{kh})$, and its covariate $\tilde{\mathbf x}_{kh}$ is lifted to the Gaussian cloud $\mathcal N(\tilde{\mathbf x}_{kh},\tilde{\bs\Sigma}_{kh})$. Subsequently, site $k$ sends $\widetilde{\mathscr D}_k=\{(n_{kh},\tilde y_{kh},\tilde{\mathbf x}_{kh},\tilde{\bs\Sigma}_{kh})\}_{h=1}^{m_k}$ to the server.

The server fits the quantile regression by gradient descent on $R_{Q_n}$, which requires the score of $R_{Q_n}$ in \eqref{eq:grad_lifted_quantile_risk_gaussian}. This score is an average over all $n$ raw observations, which the server cannot access. To construct the representatives, we require that the score of a representative equals that of its group at the current fit, so the server can compute the score of $R_{Q_n}$ from the representatives. For simplicity, we assume $m_k=1$ in this section and use $k$ as the group index, with index set $I_k$, size $n_k$, and clouds $Q_k$. Let $R_{Q_k}(\bs\theta)=n_k^{-1}\sum_{i\in I_k}\{r_iw_i+s_{\bs\beta}\phi(a_i)\}$ be the clouded risk of the group. Its score is
\begin{align}\label{eq:grad_group}
\nabla_{\bs\beta} R_{Q_k}=\frac{1}{n_k}\sum_{i\in I_k}\Bigl\{-\mathbf x_i w_i+\phi(a_i)\frac{s_{\bs\beta}\bs\beta}{\|\bs\beta\|^2}\Bigr\},\qquad
\nabla_{\beta_0} R_{Q_k}=-\frac{1}{n_k}\sum_{i\in I_k} w_i.
\end{align}

 Let $\widetilde Q_k$ denote the cloud of the representative $(n_k,\tilde y_k,\tilde{\mathbf x}_k,\tilde{\bs\Sigma}_k)$ of the group.
By the same computation as \eqref{eq:expected_check_loss}, the clouded risk of the representative is
\begin{align}\label{eq:lifted_quantile_risk_smr}
 R_{\widetilde Q_k}(\bs\theta)
 = \tilde{r}_k\,\tilde{w}_k + \tilde\sigma_k\,\phi\bigl(\tilde{a}_k\bigr),
\end{align}
where $\tilde{r}_k(\bs\theta)=\tilde{y}_k-\tilde{\mathbf{x}}_k^\top\bs\beta-\beta_0$,
$\tilde\sigma_k(\bs\beta)=\bigl(\bs\beta^\top\tilde{\bs\Sigma}_k\bs\beta\bigr)^{1/2}$,
$\tilde{a}_k(\bs\theta)=\tilde{r}_k(\bs\theta)/\tilde\sigma_k(\bs\beta)$, and
$\tilde{w}_k(\bs\theta)=\tau-\Phi\bigl(-\tilde{a}_k(\bs\theta)\bigr)$, for $\bs\beta$ with $\tilde\sigma_k(\bs\beta)>0$.
The gradient has the same two blocks as
the group score \eqref{eq:grad_group}:
\begin{align}\label{eq:grad_smr}
\nabla_{\bs\beta} R_{\widetilde Q_k}= -\tilde{\mathbf{x}}_k\,\tilde{w}_k + \phi\bigl(\tilde{a}_k\bigr)\frac{\tilde{\bs\Sigma}_k\bs\beta}{\tilde\sigma_k},\qquad
\nabla_{\beta_0} R_{\widetilde Q_k}= -\tilde{w}_k.
\end{align}

Representatives are constructed at a fixed reference value, the \textit{anchor}
$\bs\theta^\circ=(\bs\beta^\circ,\beta_0^\circ)$, which is the current fit sent by the server. We write $r_i^\circ=r_i(\bs\theta^\circ)$,
$s_{\bs\beta^\circ}=s\|\bs\beta^\circ\|$, $a_i^\circ=a_i(\bs\theta^\circ)$ and $w_i^\circ=w_i(\bs\theta^\circ)$
for the data quantities at the anchor.
Matching the representative score \eqref{eq:grad_smr} to the group score \eqref{eq:grad_group}
at the anchor, $\nabla_{\bs\theta} R_{\widetilde Q_k}(\bs\theta^\circ)=\nabla_{\bs\theta} R_{Q_k}(\bs\theta^\circ)$,
gives componentwise
\begin{align}
 \frac{1}{n_k}\sum_{i\in I_k}\left\{\phi\bigl(a_i^\circ\bigr)\frac{s_{\bs\beta^\circ}\bs\beta^\circ}{\|\bs\beta^\circ\|^2}
 -\mathbf{x}_i\, w_i^\circ\right\}
 &= \phi\bigl(\tilde{a}_k(\bs\theta^\circ)\bigr)\frac{\tilde{\bs\Sigma}_k\bs\beta^\circ}{\tilde\sigma_k(\bs\beta^\circ)}
 -\tilde{\mathbf{x}}_k\,\tilde{w}_k(\bs\theta^\circ), \label{eq:lifted_smr}\\
 \frac{1}{n_k}\sum_{i\in I_k}\Phi\bigl(a_i^\circ\bigr) &= \Phi\bigl(\tilde{a}_k(\bs\theta^\circ)\bigr).
 \label{eq:lifted_smr0}
\end{align}
We call a representative satisfying \eqref{eq:lifted_smr} and \eqref{eq:lifted_smr0} \textit{exact} at the anchor, and Proposition~\ref{prop:clouded_smr} constructs an exact representative for a group of any size.

\begin{proposition}\label{prop:clouded_smr}
Suppose $\bs\beta^\circ\neq\mathbf 0$ and $\sum_{i\in I_k} w_i^\circ\neq 0$. Then the representative given by
\begin{equation*}
\begin{aligned}
 \tilde{a}_k & = \Phi^{-1} \left(\frac{1}{n_k}\sum_{i\in I_k} \Phi(a_i^\circ)\right) \\
 \tilde{s}_k & =
 \left\{\phi\left(\Phi^{-1} \left(\frac{1}{n_k}\sum_{i\in I_k} \Phi(a_i^\circ)\right)\right)\right\}^{-1}
 \frac{1}{n_k}\sum_{i\in I_k} \phi\bigl(a_i^\circ\bigr) s_{\bs\beta^\circ}\\
 \tilde{\mathbf{x}}_k &= \left\{ \sum_{i\in I_k} w_i^\circ \right\}^{-1} \sum_{i\in I_k} w_i^\circ \mathbf{x}_i\\
 \tilde{y}_k &= \tilde{a}_k \tilde{s}_k + \tilde{\mathbf{x}}_k^\top \bs\beta^\circ+\beta_0^\circ,
\end{aligned}
\end{equation*}
and any covariance matrix $\tilde{\bs\Sigma}_k$ with $\tilde{\bs\Sigma}_k \bs\beta^\circ = (\tilde{s}_k^2/\|\bs\beta^\circ\|^2)\,\bs\beta^\circ$ satisfies the score-matching equations \eqref{eq:lifted_smr} and \eqref{eq:lifted_smr0}.
\end{proposition}
In the above construction, the covariance is not unique, since any covariance matrix with $\tilde{\bs\Sigma}_k\bs\beta^\circ=(\tilde s_k^2/\|\bs\beta^\circ\|^2)\bs\beta^\circ$ gives an exact representative.
For practical implementation, we adopt a representative covariance that preserves local second-order structure while satisfying these constraints. Let $\bar{\mathbf x}_k=n_k^{-1}\sum_{i\in I_k}\mathbf x_i$, $\mathbf C_k=n_k^{-1}\sum_{i\in I_k}(\mathbf x_i-\bar{\mathbf x}_k)(\mathbf x_i-\bar{\mathbf x}_k)^\top$, $\mathbf u=\bs\beta^\circ/\|\bs\beta^\circ\|$, and $\mathbf P_\perp=\mathbf I_p-\mathbf u\mathbf u^\top$. We construct the representative covariance as
\begin{align}\label{eq:lifted_smr_sigma}
	\tilde{\bs\Sigma}_k = \mathbf{P}_\perp \mathbf{C}_k \mathbf{P}_\perp + \frac{\tilde{s}_k^2}{\|\bs\beta^\circ\|^2} \mathbf u\mathbf u^\top.
\end{align}
This construction keeps the empirical covariance $\mathbf C_k$ of the group in the directions orthogonal to $\bs\beta^\circ$, and it satisfies the condition of Proposition~\ref{prop:clouded_smr}.

The representative covariate $\tilde{\mathbf x}_k$ in Proposition~\ref{prop:clouded_smr} is a weighted average of the covariates of the group, with the score weights $w_i^\circ$. Note that $w_i^\circ=\tau-\Phi(-a_i^\circ)$ is positive for an observation above the line $r=-\Phi^{-1}(\tau)s_{\bs\beta^\circ}$ and negative for one below it. To keep $\tilde{\mathbf x}_k$ in the convex hull of the group, we require all $w_i^\circ$ in a group to have the same sign. We thus split the observations at site $k$ into two groups, $\{i: w_i^\circ\ge0\}$ and $\{i: w_i^\circ<0\}$, and construct one representative for each group.

\subsection{The cloudqr Algorithm}\label{sec:algorithm}

Let $R_{\widetilde Q_n}=\sum_{k=1}^K\sum_{h=1}^{m_k}(n_{kh}/n)R_{\widetilde Q_{kh}}$ be the representative risk. The server computes its score, $\nabla R_{\widetilde Q_n}=\sum_{k=1}^K\sum_{h=1}^{m_k}(n_{kh}/n)\nabla R_{\widetilde Q_{kh}}$, from \eqref{eq:grad_smr}, and by Proposition~\ref{prop:clouded_smr} this score equals the score of $R_{Q_n}$ at the anchor. The server thus uses gradient descent with $\nabla R_{\widetilde Q_n}$ in place of \eqref{eq:grad_lifted_quantile_risk_gaussian}. The equality holds at the anchor only, and the difference between the two scores can grow as the iterates move away from the anchor. The representatives are therefore rebuilt at the current iterate, which becomes the new anchor, as the descent proceeds. Each rebuild requires one round of communication. To reduce the communication cost, the server conducts $g$ steps of gradient descent between two rebuilds.

To initialize cloudqr, each site sends $n_k$, $\sum_{i\in I_k}\mathbf x_i$, $\sum_{i\in I_k}\mathbf x_i\mathbf x_i^\top$, $\sum_{i\in I_k}y_i$, and $\sum_{i\in I_k}\mathbf x_iy_i$, and the server computes the pooled least-squares fit $\bs\theta^{(0)}$ and broadcasts it to the sites.

For round $t=0,1,2,\ldots$, cloudqr proceeds as follows.
\begin{enumerate}
\item Each site splits its observations into two groups by the sign of $w_i$ at $\bs\theta^{(t)}$, builds one clouded representative per group by Proposition~\ref{prop:clouded_smr} and \eqref{eq:lifted_smr_sigma} with the anchor $\bs\theta^\circ=\bs\theta^{(t)}$, and sends the two representatives to the server.
\item Starting from $\bs\theta^{(t)}$, the server conducts $g$ steps of gradient descent, $\bs\theta\leftarrow\bs\theta-\alpha\nabla R_{\widetilde Q_n}(\bs\theta)$, with the step size $\alpha=1/L$, where $L$ is the constant in Theorem~\ref{thm:clouded_convergence}, and the final iterate is $\bs\theta^{(t+1)}$. In our implementation, $\alpha$ is halved until the representative risk decreases.
\item The server broadcasts $\bs\theta^{(t+1)}$ and stops if $\|\bs\theta^{(t+1)}-\bs\theta^{(t)}\|$ is below a tolerance.
\end{enumerate}

In our implementation, we begin with $g=20$. If $(\bs\theta^{(t+1)}-\bs\theta^{(t)})^\top(\bs\theta^{(t)}-\bs\theta^{(t-1)})<0$ for some $t\ge1$, the iterates oscillate, and $g$ is halved for the following rounds, $g\leftarrow\max(1,\lfloor g/2\rfloor)$. If the representatives of site $k$ for round $t$ have not arrived, the server uses the ones that site $k$ built at an earlier fit $\bs\theta^{(t-\ell)}$ with $\ell\ge1$ until they arrive.

We standardize the covariates with the pooled mean $\bar{\mathbf x}$ and covariance $\mathbf S$, computed from the sums of the initialization, by replacing $\mathbf x_i$ with $\mathbf S^{-1/2}(\mathbf x_i-\bar{\mathbf x})$, and run cloudqr on the standardized data with $s=n^{-1/3}$, the order of the bandwidth that minimizes the mean squared error of the smoothed estimator \citep{fernandes2021smoothing}. For $n$ beyond $10{,}000$ we keep $s$ at its value for $n=10{,}000$, that is, $s=\max(n^{-1/3},10{,}000^{-1/3})$, so that the step size of the server, which is proportional to $s$, does not vanish for very large $n$. The fit $(\bs\beta,\beta_0)$ on the standardized data is mapped back to the original scale by $\bs\beta\leftarrow\mathbf S^{-1/2}\bs\beta$ and $\beta_0\leftarrow\beta_0-\bar{\mathbf x}^\top\bs\beta$. A site sends at most $2\{2+p+p(p+1)/2\}$ numbers in a round, and the server evaluates a score of at most $2K$ terms $g$ times.
\section{Theoretical Studies}\label{sec:theory}

In this section we compare the minimizers of the clouded and representative risks and establish convergence results for cloudqr. The key step is to bound the difference between the two scores away from the anchors.

We first consider a group $I_k$ of size $n_k$ with anchor $\bs\theta^\circ=(\bs\beta^\circ,\beta_0^\circ)$. Define
\begin{equation*}
\begin{aligned}
\bar r_k=\frac{1}{n_k}\sum_{i\in I_k}r_i^\circ,
\qquad
v_k=\max_{i\in I_k}|r_i^\circ-\bar r_k|,
\qquad
\bar a_k=\frac{\bar r_k}{s_{\bs\beta^\circ}},
\qquad
\Delta_k=\max_{i\in I_k}\|\mathbf x_i-\tilde{\mathbf x}_k\|.
\end{aligned}
\end{equation*}
The quantities $v_k$ and $\Delta_k$ measure the residual and covariate dispersion within the group. Assume that $\bar a_k\in\mathcal A$ and $v_k/s_{\bs\beta^\circ}\le\rho$ for a fixed compact interval $\mathcal A$ and a constant $\rho>0$. The first lemma bounds the difference between the residual of the representative and each residual in its group.

\begin{lemma}\label{lem:residual_control}
Let $\bs\beta^\circ\neq\mathbf 0$, and let $\tilde a_k$ and $\tilde s_k$ be as in Proposition~\ref{prop:clouded_smr}. Under the preceding assumptions, there exists a constant $C_{\mathcal A}>0$, depending only on $\mathcal A$ and $\rho$, such that
\begin{equation*}
\begin{aligned}
|a_i^\circ s_{\bs\beta^\circ}-\tilde a_k\tilde s_k|
\le v_k+C_{\mathcal A}\frac{v_k^2}{s_{\bs\beta^\circ}},
\qquad i\in I_k.
\end{aligned}
\end{equation*}
In particular, if $v_k=o(s_{\bs\beta^\circ})$, then $|a_i^\circ s_{\bs\beta^\circ}-\tilde a_k\tilde s_k|=O(v_k)$.
\end{lemma}

Lemma~\ref{lem:residual_control} shows that the representative preserves the group residuals to order $v_k$ when their dispersion is small relative to the cloud scale $s_{\bs\beta^\circ}$. This residual bound is used to compare the Hessians of the two risks.

For a compact parameter set $\Theta_0$, write the anchor of group $k$ as $\bs\theta_k^\circ=(\bs\beta_k^\circ,\beta_{0k}^\circ)$. For a matrix, $\|\cdot\|$ denotes the operator norm.

\begin{lemma}\label{lem:hessian_gap_node}
Let $I_k$ be a sign group with representative covariance \eqref{eq:lifted_smr_sigma} constructed at $\bs\theta_k^\circ$. Let $\psi_k(\bs\beta)$ be the angle between $\bs\beta$ and $\bs\beta_k^\circ$. Assume that
\begin{equation*}
\begin{aligned}
b_0=\inf_{\bs\theta\in\Theta_0}\|\bs\beta\|>0,
\qquad
a_i(\bs\theta),\ \tilde a_k(\bs\theta)\in\mathcal A,
\qquad
|\cos\psi_k(\bs\beta)|\ge c_0>0
\end{aligned}
\end{equation*}
for every $\bs\theta\in\Theta_0$ and $i\in I_k$. Then, for $0<s\le1$ and $\bs\theta\in\Theta_0$,
\begin{equation}\label{eq:hessian_gap_bound}
\begin{aligned}
&\left\|\nabla_{\bs\theta}^2R_{Q_k}(\bs\theta)-\nabla_{\bs\theta}^2R_{\widetilde Q_k}(\bs\theta)\right\|\\
&\quad\le\frac{C}{s_{\bs\beta_k^\circ}}\left\{\Delta_k+s+\frac{v_k}{s_{\bs\beta_k^\circ}}+\frac{v_k^2}{s_{\bs\beta_k^\circ}^2}
+\frac{\Delta_k}{s}\|\bs\beta-\bs\beta_k^\circ\|
+\left(1+\frac{\Delta_k^2}{s^2}\right)\sin^2\psi_k(\bs\beta)\right\}.
\end{aligned}
\end{equation}
Here $C>0$ depends only on $\mathcal A$, $\rho$, $c_0$, $\max_{i\in I_k}\|\mathbf x_i\|$, $b_0$, and $\|\bs\beta_k^\circ\|$.
\end{lemma}

The last two terms in the bound vanish at the anchor. Since the scores agree there, integrating the Hessian bound gives a bound on the score error over $\Theta_0$. Define
\begin{equation*}
\begin{aligned}
\mathcal E_k(\Theta_0)
&=\sup_{\bs\theta\in\Theta_0}\left\|\nabla_{\bs\theta}R_{Q_k}(\bs\theta)-\nabla_{\bs\theta}R_{\widetilde Q_k}(\bs\theta)\right\|,\\
d_k(\Theta_0)
&=\sup_{\bs\theta\in\Theta_0}\|\bs\theta-\bs\theta_k^\circ\|,
\end{aligned}
\end{equation*}
the maximum score error and the maximum distance from the anchor over $\Theta_0$.

\begin{lemma}\label{lem:local_grad_fide}
Under the conditions of Lemma~\ref{lem:hessian_gap_node}, suppose also that $\Theta_0$ is convex and contains $\bs\theta_k^\circ$. With $b_0$ and $C$ as in that lemma, define
\begin{equation*}
\begin{aligned}
\Lambda_k=\frac{C}{s_{\bs\beta_k^\circ}}\left\{\Delta_k+s+\frac{v_k}{s_{\bs\beta_k^\circ}}+\frac{v_k^2}{s_{\bs\beta_k^\circ}^2}
+\frac{\Delta_k}{2s}\,d_k(\Theta_0)
+\left(1+\frac{\Delta_k^2}{s^2}\right)\frac{d_k(\Theta_0)^2}{3b_0^2}\right\}.
\end{aligned}
\end{equation*}
Then
\begin{equation*}
\begin{aligned}
\|\nabla_{\bs\theta}R_{Q_k}(\bs\theta)-\nabla_{\bs\theta}R_{\widetilde Q_k}(\bs\theta)\|
\le\Lambda_k\|\bs\theta-\bs\theta_k^\circ\|,
\qquad\bs\theta\in\Theta_0,
\end{aligned}
\end{equation*}
and consequently $\mathcal E_k(\Theta_0)\le\Lambda_k d_k(\Theta_0)$.
\end{lemma}

Lemma~\ref{lem:local_grad_fide} shows how the score error depends on both the variation within a group and the distance from its anchor. Strong convexity of the clouded risk then gives a bound on the difference between the minimizers.

\begin{theorem}\label{thm:param_consistency}
Let $\hat{\bs\theta}$ and $\tilde{\bs\theta}$ minimize $R_{Q_n}$ and $R_{\widetilde Q_n}$, respectively, where the representative of group $k$ is built at $\bs\theta_k^\circ$ by Proposition~\ref{prop:clouded_smr}, with $\sum_{i\in I_k}w_i^\circ\neq0$; when a site holds several groups, the index $k$ runs over the groups, each with its own $n_k$, $\Delta_k$, and $v_k$. Suppose that the minimizers and all anchors belong to a convex region $\Theta_0$ on which the conditions of Lemma~\ref{lem:hessian_gap_node} hold for every group and $R_{Q_n}$ is $\mu$-strongly convex.

Let $\Lambda_k$ be the coefficient in Lemma~\ref{lem:local_grad_fide}, using the largest of the group constants $C$ in Lemma~\ref{lem:hessian_gap_node}, and set $\Lambda=\sum_{k=1}^K(n_k/n)\Lambda_k$. If $\Lambda<\mu$, then
\begin{equation*}
\begin{aligned}
\|\tilde{\bs\theta}-\hat{\bs\theta}\|
\le\frac{1}{\mu-\Lambda}\sum_{k=1}^K\frac{n_k}{n}\,\Lambda_k\,\|\hat{\bs\theta}-\bs\theta_k^\circ\|.
\end{aligned}
\end{equation*}
\end{theorem}

Theorem~\ref{thm:param_consistency} shows that $\tilde{\bs\theta}=\hat{\bs\theta}$ when every anchor equals $\hat{\bs\theta}$. For other anchors, the error depends on their distances to $\hat{\bs\theta}$ and on the ratio $\Lambda/\mu$.

To study cloudqr, we allow the anchors to change during the iteration. In the following theorem, $t$ counts individual server steps and $k$ indexes sites.

\begin{theorem}\label{thm:clouded_convergence}
At step $t$, let the representatives of site $k$ be anchored at $\bs\theta^{(a_k(t))}$, where $a_k(t)\le t$, and let $R_{\widetilde Q_n}^{(t)}$ be the corresponding representative risk. Consider the update
\begin{equation*}
\begin{aligned}
\bs\theta^{(t+1)}=\bs\theta^{(t)}-\alpha\nabla R_{\widetilde Q_n}^{(t)}(\bs\theta^{(t)}).
\end{aligned}
\end{equation*}
Suppose that all iterates and anchors lie in a convex region $\Theta_0$ containing the minimizer $\hat{\bs\theta}$ of $R_{Q_n}$. Assume that $R_{Q_n}$ is twice continuously differentiable on a neighborhood of $\Theta_0$ and
\begin{equation*}
\mu\mathbf I\preceq\nabla^2R_{Q_n}(\bs\theta)\preceq L\mathbf I
\end{equation*}
for all $\bs\theta\in\Theta_0$ and some constants $0<\mu\le L$.
For a fixed step size $0<\alpha\le1/L$, define the score error
\begin{equation*}
\begin{aligned}
e^{(t)}=\nabla R_{\widetilde Q_n}^{(t)}(\bs\theta^{(t)})-\nabla R_{Q_n}(\bs\theta^{(t)}).
\end{aligned}
\end{equation*}
Then, for every $T\ge1$,
\begin{equation*}
\begin{aligned}
\|\bs\theta^{(T)}-\hat{\bs\theta}\|
\le(1-\alpha\mu)^T\|\bs\theta^{(0)}-\hat{\bs\theta}\|
+\frac{\max_{t<T}\|e^{(t)}\|}{\mu}.
\end{aligned}
\end{equation*}
\end{theorem}

Theorem~\ref{thm:clouded_convergence} gives geometric convergence to $\hat{\bs\theta}$ when $e^{(t)}=\mathbf 0$ at every step. For uniformly bounded score errors, it gives $\limsup_{T\to\infty}\|\bs\theta^{(T)}-\hat{\bs\theta}\|\le\sup_t\|e^{(t)}\|/\mu$.

\begin{remark}\label{rem:schedules}
Under the conditions of Theorem~\ref{thm:clouded_convergence}, suppose also that Lemma~\ref{lem:hessian_gap_node} holds uniformly over all groups and anchors used by the server. For site $k$, let $\Lambda_k$ bound the weighted sum of its group coefficients in Lemma~\ref{lem:local_grad_fide}, with weights $n_{kh}/n_k$, uniformly over iterations. Then
\begin{equation*}
\begin{aligned}
\|e^{(t)}\|\le\sum_{k=1}^K\frac{n_k}{n}\,\Lambda_k\,\bigl\|\bs\theta^{(t)}-\bs\theta^{(a_k(t))}\bigr\|.
\end{aligned}
\end{equation*}
We distinguish three cases.
\begin{enumerate}
\item[(a)] \textit{Representatives rebuilt at every step.} If $a_k(t)=t$ for all $k$, Proposition~\ref{prop:clouded_smr} gives $e^{(t)}=\mathbf 0$. The iterates converge geometrically to $\hat{\bs\theta}$.
\item[(b)] \textit{Representatives reused for a bounded number of steps.} Suppose that $t-a_k(t)\le m$ for all $k$ and $t$, where $m$ is finite. Since
\begin{equation*}
\begin{aligned}
\|\bs\theta^{(t)}-\bs\theta^{(a_k(t))}\|
\le m\max_{\max(0,t-m)\le u<t}\|\bs\theta^{(u+1)}-\bs\theta^{(u)}\|,
\end{aligned}
\end{equation*}
the error $e^{(t)}$ tends to zero if the step lengths do. Applying Theorem~\ref{thm:clouded_convergence} from a sufficiently late iterate then gives convergence to $\hat{\bs\theta}$. In particular, any convergent iteration has this limit.
\item[(c)] \textit{Representatives held fixed.} Suppose that $a_k(t)=0$ for all $k$ and $t$. If also $\alpha\le1/\widetilde L$, where $\widetilde L$ bounds the Hessian of the fixed representative risk on $\Theta_0$, the iteration converges to a minimizer $\tilde{\bs\theta}$ of that risk. Theorem~\ref{thm:param_consistency} bounds $\|\tilde{\bs\theta}-\hat{\bs\theta}\|$ under its conditions.
\end{enumerate}
In cloudqr, a representative that is $D$ rounds old is at most $(D+1)g_0$ server steps old, where $g_0$ is the initial number of steps per round. This follows because the halving rule in Section~\ref{sec:algorithm} can only reduce the number of steps per round.
\end{remark}

\section{Simulation Studies}\label{sec:simu}

In this section, we compare the performance of cloudqr with existing distributed methods, using the pooled fit as the reference. We mainly focus on three practical situations: (1) the sites are heterogeneous, (2) the sites are small, and (3) some sites reply late.

We generated $K$ sites with $n_k$ observations per site and $p=20$ covariates, giving $n=Kn_k$ observations in total. For each site $k$, we drew a center $\bs\mu_k$ whose coordinates were independent and uniform on $(-1,1)$. We then generated $\mathbf x_i=\bs\mu_k+\sigma\mathbf z_i$ for $i\in I_k$, where $\mathbf z_i\sim \mathcal N_p(\mathbf 0,\bs\Sigma_e)$, and $\bs\Sigma_e$ has unit diagonal and off-diagonal entries $1/2$. The parameter $\sigma$ controls the spread of the covariates around the site center. The responses were generated from the location-scale model $y_i=\mathbf x_i^\top\bs\beta^*+(1+\mathbf x_i^\top\bs\gamma)\varepsilon_i$ with standard normal errors $\varepsilon_i$. We used $\bs\beta^*\sim \mathcal N_p(\mathbf 0,\mathbf I_p)$ and chose a random direction for $\bs\gamma$, scaled so that $\max_i|\mathbf x_i^\top\bs\gamma|=0.9$. In the baseline setting, we used $n_k=200$, $K=50$, and $\sigma=0.5$, giving $n=10{,}000$. We also generated a test set of the same size from the same sites.

We compared cloudqr with Avg \citep{volgushev2019distributed}, which averages the quantile regression fits from each site. We computed the pooled fit, $\hat{\bs\theta}_{\mathrm{QR}}=(\hat{\bs\beta}_{\mathrm{QR}},\hat\beta_{0,\mathrm{QR}})$ in \eqref{eq:pooled_qr}, and the fit of each site in Avg with the R package quantreg \citep{koenker2025quantreg}. We included moment-matching representatives, denoted by MMR, which lift the point representative of each site \citep{li2022smr,li2026asmr}, its mean covariate and mean response, to one Gaussian cloud with covariance $\mathbf C_k+s^2\mathbf I_p$, with $\mathbf C_k$ as in \eqref{eq:lifted_smr_sigma}. Each site sends its cloud once, and the server fits the model on these clouds without refreshing them. We then applied FedAvg and FedSGD \citep{mcmahan2017communication}, two popular federated learning methods, to the Gaussian-smoothed check loss of \citet{he2023smoothed} with its default bandwidth; FedAvg takes $E=10$ local gradient steps per round with the step size $1/\max_k L_k$, where $L_k$ is the Lipschitz constant of the gradient of the smoothed loss at site $k$, and FedSGD is the special case of FedAvg with $E=1$. We also considered distributed quantile regression via convolution smoothing \citep[DQR-CS;][]{tan2022communication}, with the largest site as the master and Avg as the initial fit.

For a fit $\hat{\bs\theta}=(\hat{\bs\beta},\hat{\beta}_0)$ at level $\tau$, we computed the ratio of its check loss to that of the pooled fit on the test set $\{(y_i^*,\mathbf x_i^*)\}_{i=1}^n$,
\begin{equation}\label{eq:simu_crit}
\begin{aligned}
\frac{\sum_{i=1}^n\rho_\tau(y_i^*-\mathbf x_i^{*\top}\hat{\bs\beta}-\hat{\beta}_0)}{\sum_{i=1}^n\rho_\tau(y_i^*-\mathbf x_i^{*\top}\hat{\bs\beta}_{\mathrm{QR}}-\hat\beta_{0,\mathrm{QR}})}.
\end{aligned}
\end{equation}
We report the averages over 50 runs at $\tau=0.5$ and $0.9$. Each iterative method ran at most 200 rounds of communication between the server and the sites.

\begin{table}[t]
\spacingset{1}
\caption{Heterogeneous sites. The ratio \eqref{eq:simu_crit} of the test check loss to that of the pooled fit, averaged over 50 independent runs. The spread $\sigma$ of a site around its center varies, and a smaller $\sigma$ gives more heterogeneous sites. The standard errors are given in parentheses. The lowest ratio in each row is in italics.}
\label{tab:sim_hetero}
\ra{0.9}
\makeatletter\let\input\@@input\makeatother
\centering\small
\begin{tabular}{lrrrrrr}
\toprule
$\sigma$ & cloudqr & Avg & MMR & FedAvg & FedSGD & DQR-CS \\
\midrule
\multicolumn{7}{c}{$\tau=0.5$} \\
0.5 &  \textit{1.000}{\scriptsize\,(0.000)} &  1.009{\scriptsize\,(0.001)} &  1.957{\scriptsize\,(0.098)} &  1.003{\scriptsize\,(0.001)} &  2.392{\scriptsize\,(0.060)} &  5.401{\scriptsize\,(1.185)} \\
0.25 &  \textit{1.000}{\scriptsize\,(0.000)} &  1.040{\scriptsize\,(0.004)} &  1.243{\scriptsize\,(0.034)} &  1.019{\scriptsize\,(0.003)} &  2.232{\scriptsize\,(0.054)} &  2.969{\scriptsize\,(0.917)} \\
0.1 &  \textit{1.000}{\scriptsize\,(0.000)} &  1.239{\scriptsize\,(0.022)} &  1.020{\scriptsize\,(0.003)} &  1.032{\scriptsize\,(0.003)} &  2.156{\scriptsize\,(0.051)} &  1.239{\scriptsize\,(0.022)} \\
0.05 &  \textit{1.000}{\scriptsize\,(0.000)} &  1.763{\scriptsize\,(0.062)} &  1.001{\scriptsize\,(0.000)} &  1.035{\scriptsize\,(0.004)} &  2.142{\scriptsize\,(0.050)} &  1.763{\scriptsize\,(0.062)} \\
\midrule
\multicolumn{7}{c}{$\tau=0.9$} \\
0.5 &  \textit{1.000}{\scriptsize\,(0.000)} &  1.022{\scriptsize\,(0.003)} &  2.319{\scriptsize\,(0.148)} &  1.113{\scriptsize\,(0.013)} &  3.281{\scriptsize\,(0.103)} &  5.250{\scriptsize\,(1.456)} \\
0.25 &  \textit{1.000}{\scriptsize\,(0.000)} &  1.086{\scriptsize\,(0.011)} &  1.447{\scriptsize\,(0.044)} &  1.168{\scriptsize\,(0.014)} &  2.965{\scriptsize\,(0.089)} &  2.241{\scriptsize\,(0.895)} \\
0.1 &  \textit{1.000}{\scriptsize\,(0.000)} &  1.605{\scriptsize\,(0.088)} &  1.470{\scriptsize\,(0.015)} &  1.190{\scriptsize\,(0.014)} &  2.818{\scriptsize\,(0.081)} &  1.605{\scriptsize\,(0.088)} \\
0.05 &  \textit{1.000}{\scriptsize\,(0.000)} &  3.168{\scriptsize\,(0.294)} &  1.693{\scriptsize\,(0.016)} &  1.196{\scriptsize\,(0.014)} &  2.792{\scriptsize\,(0.080)} &  3.168{\scriptsize\,(0.294)} \\

\bottomrule
\end{tabular}
\end{table}

\paragraph{Heterogeneous sites.} We first investigated the performance of the methods when the sites are heterogeneous. We varied $\sigma$ over $0.5$, $0.25$, $0.1$, and $0.05$ while keeping $n_k=200$ and $K=50$. With the centers uniform on $(-1,1)^p$, the sites overlap at $\sigma=0.5$ and are well separated at $\sigma=0.05$. A smaller $\sigma$ thus gives more heterogeneous sites, because the covariates of each site stay closer to its own center and overlap less with those of the other sites.

From Table~\ref{tab:sim_hetero} we observe that cloudqr matches the pooled fit in every row, with a ratio of 1.000. The loss of Avg increases as $\sigma$ decreases from 0.5 to 0.05, and its ratio rises from 1.009 to 1.763 at $\tau=0.5$ and from 1.022 to 3.168 at $\tau=0.9$. DQR-CS starts from Avg and refines the fit at the master site in each round. At $\sigma=0.1$ and $0.05$, this refinement did not converge in the first round, so DQR-CS returned Avg and the two methods have the same ratios. We also discover that MMR improves at $\tau=0.5$ as $\sigma$ decreases, but its loss at $\tau=0.9$ remains high. FedAvg and FedSGD both have higher losses as $\tau$ increases.

\begin{table}[t]
\spacingset{1}
\caption{Small sites. The ratio \eqref{eq:simu_crit} of the test check loss to that of the pooled fit, averaged over 50 independent runs, with $n=10{,}000$ observations split into $K$ sites of $n_k$ observations. The standard errors are given in parentheses. Cases are marked as a dash when the method does not produce an estimate. The lowest ratio in each row is in italics.}
\label{tab:sim_small}
\ra{0.9}
\makeatletter\let\input\@@input\makeatother
\centering\small
\begin{tabular}{llrrrrrr}
\toprule
$n_k$ & $K$ & cloudqr & Avg & MMR & FedAvg & FedSGD & DQR-CS \\
\midrule
\multicolumn{8}{c}{$\tau=0.5$} \\
200 & 50 &  \textit{1.000}{\scriptsize\,(0.000)} &  1.009{\scriptsize\,(0.001)} &  1.957{\scriptsize\,(0.098)} &  1.003{\scriptsize\,(0.001)} &  2.392{\scriptsize\,(0.060)} &  5.401{\scriptsize\,(1.185)} \\
50 & 200 &  \textit{1.000}{\scriptsize\,(0.000)} &  1.020{\scriptsize\,(0.002)} &  1.784{\scriptsize\,(0.110)} &  1.001{\scriptsize\,(0.000)} &  2.514{\scriptsize\,(0.070)} &  4.463{\scriptsize\,(0.706)} \\
5 & 2,000 &  \textit{1.000}{\scriptsize\,(0.000)} & -- &  1.632{\scriptsize\,(0.080)} &  1.038{\scriptsize\,(0.007)} &  2.895{\scriptsize\,(0.075)} & -- \\
\midrule
\multicolumn{8}{c}{$\tau=0.9$} \\
200 & 50 &  \textit{1.000}{\scriptsize\,(0.000)} &  1.022{\scriptsize\,(0.003)} &  2.319{\scriptsize\,(0.148)} &  1.113{\scriptsize\,(0.013)} &  3.281{\scriptsize\,(0.103)} &  5.250{\scriptsize\,(1.456)} \\
50 & 200 &  \textit{1.000}{\scriptsize\,(0.000)} &  1.155{\scriptsize\,(0.013)} &  2.004{\scriptsize\,(0.157)} &  1.112{\scriptsize\,(0.014)} &  3.634{\scriptsize\,(0.134)} &  1.155{\scriptsize\,(0.013)} \\
5 & 2,000 &  \textit{1.000}{\scriptsize\,(0.000)} & -- &  1.795{\scriptsize\,(0.094)} &  1.360{\scriptsize\,(0.027)} &  5.026{\scriptsize\,(0.163)} & -- \\

\bottomrule
\end{tabular}
\end{table}

\paragraph{Small sites.} We kept $n=10{,}000$ and $\sigma=0.5$ and used $K=50$, $200$, and $2{,}000$ sites, with $n_k=200$, $50$, and $5$ observations per site, respectively. 

We have several observations from Table~\ref{tab:sim_small}. First, cloudqr matches the pooled fit at both quantiles for every site size. In contrast, the ratio of FedAvg is 1.360 at $\tau=0.9$ with $n_k=5$ and $K=2{,}000$. Second, Avg and DQR-CS return no estimate at $n_k=5$. This is because five observations cannot determine a quantile regression with $21$ coefficients. Third, MMR improves as the sites become smaller, but its check loss remains well above that of cloudqr.

\begin{table}[t]
\spacingset{1}
\caption{Late replies. The ratio \eqref{eq:simu_crit} of the test check loss to that of the pooled fit, averaged over 50 independent runs, with half of the sites replying $\ell$ rounds late and the other half on time. A late site works from the fit that the central server broadcast $\ell$ rounds earlier. The standard errors are given in parentheses. The lowest ratio in each row is in italics.}
\label{tab:sim_late}
\ra{0.9}
\makeatletter\let\input\@@input\makeatother
\centering\small
\begin{tabular}{lrrrrr}
\toprule
$\ell$ & cloudqr & FedAvg & FedSGD & FedAsync & FedBuff \\
\midrule
\multicolumn{6}{c}{$\tau=0.5$} \\
0 &  \textit{1.000}{\scriptsize\,(0.000)} &  1.003{\scriptsize\,(0.001)} &  2.392{\scriptsize\,(0.060)} &  1.005{\scriptsize\,(0.001)} &  1.000{\scriptsize\,(0.000)} \\
10 &  \textit{1.000}{\scriptsize\,(0.000)} &  1.956{\scriptsize\,(0.048)} &  3.279{\scriptsize\,(0.087)} &  1.426{\scriptsize\,(0.032)} &  1.001{\scriptsize\,(0.000)} \\
50 &  \textit{1.000}{\scriptsize\,(0.000)} &  2.976{\scriptsize\,(0.075)} &  3.482{\scriptsize\,(0.095)} &  2.034{\scriptsize\,(0.052)} &  1.001{\scriptsize\,(0.000)} \\
\midrule
\multicolumn{6}{c}{$\tau=0.9$} \\
0 &  \textit{1.000}{\scriptsize\,(0.000)} &  1.113{\scriptsize\,(0.013)} &  3.281{\scriptsize\,(0.103)} &  1.136{\scriptsize\,(0.014)} &  1.003{\scriptsize\,(0.001)} \\
10 &  \textit{1.000}{\scriptsize\,(0.000)} &  2.439{\scriptsize\,(0.070)} &  6.627{\scriptsize\,(0.195)} &  1.804{\scriptsize\,(0.045)} &  1.039{\scriptsize\,(0.006)} \\
50 &  \textit{1.063}{\scriptsize\,(0.004)} &  5.425{\scriptsize\,(0.157)} &  7.565{\scriptsize\,(0.223)} &  2.702{\scriptsize\,(0.081)} &  1.090{\scriptsize\,(0.011)} \\

\bottomrule
\end{tabular}
\end{table}

\paragraph{Late replies.} We now consider a situation where half of the sites replied $\ell$ rounds late, with $\ell=10$ and $50$, and the other half replied on time. A late site works from the fit broadcast $\ell$ rounds earlier, as described in Section~\ref{sec:algorithm}, and we used this delay convention for every iterative method.

Avg and MMR communicate once and are unaffected by delays, and DQR-CS stops after its first round in these settings. We therefore excluded these three methods and added FedAsync \citep{xie2020asynchronousfederatedoptimization} and FedBuff \citep{nguyen2022federated}, two popular asynchronous federated learning methods, which update the server fit as replies arrive and downweight a reply by its staleness. Both methods used the smoothed check loss and $E=10$ local steps of FedAvg, and a late site works from the fit it received $\ell$ rounds earlier.

From Table~\ref{tab:sim_late} we observe that late replies have little effect on cloudqr at $\tau=0.5$. FedBuff is also competitive at $\tau=0.5$, and both methods match the pooled fit when $\ell=0$. At $\tau=0.9$, cloudqr loses accuracy with the longer delay. With $\ell=50$ and $\tau=0.9$, its ratio is 1.063, whereas that of FedBuff is 1.090. FedAvg, FedSGD, and FedAsync incur much larger losses as the delay increases.

\paragraph{Summary.}  The simulations show that cloudqr is highly competitive in the standard setting. The advantage over Avg and FedAvg increases as the sites become more heterogeneous, and cloudqr also remains accurate when each site has fewer observations than regression coefficients, where Avg and DQR-CS fail to work. We further observe that cloudqr compares favorably with the competing methods when some sites reply late.

\section{Real Data Analysis}\label{sec:data}

We applied cloudqr to hospital discharge records and small business loans, which feature strong heterogeneity across sites, and many of the sites are small.

The first data set came from the Statewide Planning and Research Cooperative System (SPARCS) of \citet{nysdoh2024sparcs}. We used a random sample of 1.5 million inpatient discharges in 2022 and treated each hospital as a site. The response was length of stay in days, and the 23 covariates described patient demographics, clinical conditions, and care. A hospital has a fixed number of beds, and when long stays are not anticipated, patients who need urgent admission wait for a bed and their care is delayed. We used $\tau=0.9$ to study long stays.

The second data set records loans under the 7(a) program, by which the U.S. Small Business Administration (SBA) guarantees part of a lender's loan to a small business \citep{sba2026foia}. We used 933,745 loans approved in fiscal years 2010 to 2026, with each lender treated as a site. The response was the initial interest rate in percent, and the 22 covariates described the loan, the business, and the approval period. Here a fit at $\tau=0.9$ can give the SBA a reference for monitoring high interest rates among comparable loans.

In these two applications, patient populations and lending practices vary across institutions, and site sizes differ greatly. The share of patients aged 70 or older is below 25\% at a quarter of the 205 hospitals and above 44\% at another quarter, and the share of loans to new businesses is below 18\% at a quarter of the 3,148 lenders and above 50\% at another quarter. The median hospital has 5,224 discharges, and 46 hospitals have fewer than 1,000. The median lender has 17 loans, and 68\% of the lenders have fewer than 50, whereas the largest lender has 82,259.

We used 50 random splits, each with 80\% of the observations for training and 20\% for testing. We kept every site with at least one training observation, so many lenders have fewer observations than coefficients. We standardized the covariates and response using the training data and used the methods and settings of Section~\ref{sec:simu}. We excluded Avg, because it needs a quantile regression fit at every site, and the design matrix is rank deficient at 165 of the 205 hospitals and 2,780 of the 3,148 lenders. Most lenders have fewer loans than coefficients, and at most hospitals a binary covariate takes a single value, such as the indicator of a trauma admission at a hospital that admits no trauma patients. DQR-CS used the largest site with a full-rank design as the master and started from the average of the fits at the sites with a full-rank design. We report the test check loss ratio \eqref{eq:simu_crit} at $\tau=0.5$ and $0.9$ after 200 rounds, averaged over the splits.

\begin{table}[t]
\spacingset{1}
\caption{Real data analysis. The ratio \eqref{eq:simu_crit} of the test check loss to that of the pooled fit, averaged over 50 random splits. The standard errors are given in parentheses. Cases are marked as a dash when the method diverges in some splits. The lowest ratio in each row is in italics.}
\label{tab:rd_loss}
\ra{0.9}
\makeatletter\let\input\@@input\makeatother
\centering\small\setlength{\tabcolsep}{3.5pt}
\begin{tabular}{lrrcrrrrr}
\toprule
data & $n$ & $K$ & $\tau$ & cloudqr & MMR & FedAvg & FedSGD & DQR-CS \\
\midrule
SPARCS & 1,500,000 & 205 & 0.5 & \textit{1.001}{\scriptsize\,(0.000)} & 1.298{\scriptsize\,(0.000)} & 1.003{\scriptsize\,(0.000)} & 1.056{\scriptsize\,(0.000)} & 1.007{\scriptsize\,(0.000)} \\
 & &  & 0.9 & \textit{1.000}{\scriptsize\,(0.000)} & 1.429{\scriptsize\,(0.000)} & 1.032{\scriptsize\,(0.000)} & 1.311{\scriptsize\,(0.000)} & 1.020{\scriptsize\,(0.000)} \\
\midrule
SBA & 933,745 & 3,148 & 0.5 & \textit{1.000}{\scriptsize\,(0.000)} & 1.635{\scriptsize\,(0.000)} & 1.097{\scriptsize\,(0.002)} & 1.724{\scriptsize\,(0.002)} & 1.035{\scriptsize\,(0.000)} \\
 & &  & 0.9 & \textit{1.000}{\scriptsize\,(0.000)} & 1.875{\scriptsize\,(0.001)} & 1.371{\scriptsize\,(0.007)} & 3.605{\scriptsize\,(0.007)} & -- \\

\bottomrule
\end{tabular}
\end{table}

From Table~\ref{tab:rd_loss} we observe that cloudqr closely matches the pooled fit and has the lowest ratio in every row. On SBA, where most lenders are small, the ratio of FedAvg is 1.097 at $\tau=0.5$ and 1.371 at $\tau=0.9$, whereas cloudqr stays at 1.000. DQR-CS diverges in some splits on SBA at $\tau=0.9$.

\section{Discussion}\label{sec:diss}

In this paper, we have proposed cloudqr for distributed quantile regression. The method lifts observations to Gaussian clouds and summarizes each group by a clouded representative that matches its score at the anchor. We have established approximation bounds and convergence guarantees for the server iteration. Our numerical studies show that cloudqr is competitive with existing distributed methods and can accommodate heterogeneous sites, small sites, and late replies. We have implemented our method in the R package cloudqr.

The clouded risk can also be extended to generalized linear models by averaging the corresponding loss over Gaussian clouds. Clouded representatives can then be constructed by matching the resulting group scores. Developing the corresponding algorithms and establishing their convergence properties would be useful future work.

\bibliographystyle{jasa}
\bibliography{references}

\end{document}